\documentstyle[aps]{revtex}

\begin{document}

\twocolumn[\hsize\textwidth\columnwidth\hsize\csname@twocolumnfalse%
\endcsname

\draft

\title{
Hydrodynamic modes in a trapped Bose gas
above the Bose-Einstein transition}

\author{A. Griffin and Wen-Chin Wu\cite{wu}}
\address{Department of Physics, University of Toronto
Toronto, Ontario, M5S 1A7, Canada}
 
\author{S. Stringari}
\address{Dipartimento di Fisica, Universit\`a di Trento
 and Istituto Nazionale di Fisica della Materia,
I-3850 Povo, Italy}

\date{\today}
 
\maketitle

\begin{abstract}
We discuss the collective modes of a trapped Bose gas in the hydrodynamic
regime where atomic collisions ensure local
thermal equilibrium for the distribution function.
Starting from the conservation laws, in the linearized
limit we derive a closed
equation for the velocity fluctuations in a trapped Bose gas above
the Bose-Einstein transition temperature. Explicit solutions
for a parabolic trap are given. 
We find that the surface modes have the same dispersion
relation as the one recently obtained by Stringari
for the oscillations of the condensate at $T=0$ within the Thomas-Fermi
approximation. Results are also given
for the monopole ``breathing'' mode as well as for the
$m=0$ excitations which result from the coupling of the monopole and quadrupole
modes in an anisotropic parabolic well. 
\end{abstract}

]

\vskip 0.1 true in

\narrowtext

It is well known in discussions of uniform quantum fluids that there
are two kinds of collective modes\cite{ref1}.  There are
collisionless (zero sound) modes arising from dynamic
self-consistent mean fields.
Recently there have been several theoretical
studies\cite{ref2,ref3}
of the collective modes of a trapped Bose gas at $T=0$, where all
the atoms are in the condensate
These are all based on the Gross-Pitaevskii description corresponding to the
time-dependent Hartree theory of the
Bose condensate, which can be also
generalized to finite temperature \cite{ref4}.
 Recently the  lowest frequency
oscillations of a trapped Bose
condensate have been measured \cite{ref5a,ref5b} and found to be in excellent
agreement with the calculated $T=0$ Bogoliubov excitation
frequencies\cite{ref6,ref3}.

In addition, there can be hydrodynamic modes (such as first
and second sound in Bose superfluids) which arise when collisions
are sufficiently strong to ensure local thermodynamic equilibrium.
This letter is concerned with the hydrodynamic modes of a trapped Bose gas.
We derive a closed equation for the velocity fluctuations
in a gas which is not Bose-condensed ($T\geq T_{BEC}$).  
We exhibit exact solutions of this
equation for a parabolic well corresponding to surface,
monopole and $m=0$ coupled monopole-quadrupole modes.
The frequencies of these normal modes are shown to be
the same for a degenerate Bose gas as for a classical gas.
The surface modes have frequencies 
identical to those found by Stringari \cite{ref3} for a $T=0$ Bose
condensate, in the Thomas-Fermi approximation \cite{comment1b},
while the monopole mode has the same frequency as in a non-interacting gas.

The hydrodynamic regime is achieved when
the collisional frequency $\omega_c=1/\tau_c$
is much larger than the frequency $\omega$ of the
collective mode ($\omega\tau_c\ll 1$). For $\omega\sim\omega_0$ (the
trap frequency), this condition is very similar to requiring
that the size of the system $R$ be much larger than the collisional
mean free path $\ell\simeq 1/n\sigma$, where
$\sigma = 8\pi a^2$ is the s-wave cross section and
$n$ is the density \cite{comment1}. At $T \sim 2 T_{BEC}$, 
these conditions seem to be well satisfied in a 
recent MIT experiment \cite{ref5a}, 
while in the JILA experiment \cite{ref5b} the system
is too dilute and is in the non-interacting regime above $T_{BEC}$.

Kadanoff and Baym \cite{ref7} have given a very clear treatment of
collision-dominated hydrodynamic modes in a {\it uniform} dilute
classical gas which is easily extended to a
degenerate Bose gas in an external potential $U_0({\bf r})$.  The
collisions ensure that the perturbed distribution function produced
by a slowly varying external field $\delta U({\bf r},t)$\ is always given
by the local thermodynamic equilibrium Bose distribution function

\begin{equation}
f({\bf p},{\bf r},t) =\left(e^{\beta({\bf r},t)[({\bf p}-m{\bf
v}({\bf r},t))^2/2m-\mu({\bf r},t)]}-
1 \right)^{-1}\ .
\label{eq1}
\end{equation}
The perturbed distribution is described by the space- and
time-dependent temperature $T({\bf r},t))\equiv [k_B\beta({\bf r},t)]^{-1}$,
chemical potential $\mu({\bf r},t)$, and local velocity {\bf v}({\bf r},t).
Inserting (\ref{eq1}) into the conservation laws for the local
particle number $n({\bf r},t)$,  particle current ${\bf j}({\bf r},t)$ and
energy density $\varepsilon({\bf r},t)$, one finds equations of motion which
determine the values of $T({\bf r},t), \mu({\bf r},t)$ and ${\bf v}
({\bf r},t)$.  (This procedure is described on p.56--58 of Ref.~\cite{ref7} 
for a {\it uniform} classical gas).  We shall only be interested in  
small perturbations around the equilibrium state,

\begin{eqnarray}
T({\bf r},t)&=& T_0 +\delta T({\bf r},t)\nonumber\\
\mu({\bf r},t) &=&\mu_0({\bf r}) +\delta\mu({\bf r},t)\label{eq2}\\
{\bf v}({\bf r},t)&=&\delta{\bf v}({\bf r},t)\ .\nonumber
\end{eqnarray}
We note that in the equilibrium state of the non-uniform trapped gas,
the temperature $T_0$ is uniform and the local velocity
${\bf v}_0({\bf r})$ is zero.

Working to first order in $\delta{\bf v}({\bf r},t)$,
the conservation laws for the local perturbed quantities

\begin{eqnarray}
n({\bf r},t)&=&\int{d{\bf p}\over(2\pi)^3} f({\bf p}, {\bf
r},t)\nonumber\\
{\bf j}({\bf r},t) &=&\int{d{\bf p}\over(2\pi)^3} {{\bf p}\over m}
f({\bf p}, {\bf r},t)\label{eq3}\\
\varepsilon({\bf r},t)&=&\int{d{\bf p}\over(2\pi)^3} {p^2\over 2m}
f({\bf p},{\bf r},t)\nonumber
\end{eqnarray}
reduce to \cite{ref7}

\begin{equation}
{\partial n({\bf r},t)\over \partial t} = -{\boldmath\nabla} \cdot
[n_0({\bf r})\delta{\bf v}({\bf r},t)]
\label{eq4}
\end{equation}

\begin{equation}
mn_0({\bf r}) {\partial\delta{\bf v}({\bf r},t)\over \partial t} =
-[{\boldmath\nabla} P({\bf r},t)
+ n({\bf r},t){\boldmath\nabla} U_0({\bf r})]
\label{eq5}
\end{equation}

\begin{equation}
{\partial\varepsilon({\bf r},t)\over \partial t} =
-{\boldmath\nabla}\cdot \left[ {5\over
3}\varepsilon_0({\bf r})\delta {\bf v}({\bf r},t)\right ] -
n_0({\bf
r}) \delta{\bf v}({\bf r},t)
\cdot {\boldmath\nabla} U_0({\bf r}).
\label{eq6}
\end{equation}
The terms proportional to ${\boldmath\nabla} U_0({\bf r})$ in (\ref{eq5})
and (\ref{eq6}) show the explicit role of the trapping potential.
The equilibrium density $n_0({\bf r})$ and pressure $P_0({\bf r})$ must be
consistent with the vanishing of the r.h.s. of (\ref{eq5}) when
$\delta {\bf v}({\bf r}, t) = 0$,

\begin{equation}
{\boldmath\nabla} P_0({\bf r}) + n_0({\bf r}){\boldmath\nabla} U_0
({\bf r}) = 0.
\label{eq7}
\end{equation}
Using (\ref{eq1}) in (\ref{eq3}), the integrals over
${\bf p}$\ are exactly the same as for a uniform Bose gas in thermal
equilibrium \cite{ref8} and one may easily verify that

\begin{eqnarray}
{\bf j}({\bf r},t) &=& n_0({\bf r}) \delta{\bf v}({\bf r},t)\
\label{eq8}\\
n({\bf r},t) &=& {1\over \Lambda^3} g_{3/2}(z) \label{eq9}\\
P({\bf r},t) &=& {1\over \beta({\bf r},t)} {1\over \Lambda^3}
g_{5/2} (z) = {2\over 3} \varepsilon({\bf r},t)\ . \label{eq10}
\end{eqnarray}
Here $z({\bf r},t)=e^{\beta(r,t)\mu (r,t)}$
is the local thermodynamic equilibrium  fugacity,
the thermal de Broglie
wavelength $\Lambda = (2\pi\hbar^2/mk_BT({\bf r},t))^{1/2}$,
and $g_n(z)=\sum^\infty_{l=1} z^l/l^n$
are the well-known Bose-Einstein functions \cite{ref8}.

While collisions are crucial to enforce the local thermodynamic
equilibrium solution given by (\ref{eq1}), for a dilute gas one
can, to lowest order,
ignore the interactions in evaluating the hydrodynamic equations of
motion from the conservation laws, as we have done in the preceding
analysis. The static local equilibrium values of the thermodynamic functions
are given by (\ref{eq8})-(\ref{eq10}) by setting
$\delta{\bf v}({\bf r},t)=0$, $T({\bf r},t)=T_0$
and $z=z_0\equiv e^{\mu_0({\bf r})/k_BT_0}$, where 
$\mu_0({\bf r}) = \mu-U_0({\bf r})$ and
$\mu$ is the chemical potential.  We note that the
temperature $T_0$ is constant throughout the trap. These results
correspond to the well-known semi-classical approximation
\cite{ref9}, which is valid when the discrete energy level spacing
of the trapping potential is much less than $k_BT_0$.  
The resulting equilibrium values are consistent with the
relation given in (\ref{eq7}), as  easily verified using the well-known
identity $\partial g_n(z)/\partial z = g_{n-1}(z)/z$.

One can use (\ref{eq4})--(\ref{eq6}) to derive
a closed equation for the velocity fluctuation  $\delta {\bf v}({\bf r},t)$.
Using (\ref{eq10}), (\ref{eq6}) can be rewritten as

\begin{equation}
{\partial  P({\bf r},t) \over \partial t} = -{5\over 3}
{\boldmath\nabla}\cdot [P_0({\bf r})\delta {\bf v}({\bf r},t)]
-{2\over 3} n_0({\bf r}) \delta{\bf v}({\bf r},t) \cdot
{\boldmath\nabla} U_0({\bf r}).
\label{eq120}
\end{equation}
Taking the time derivative of (\ref{eq5}) and using (\ref{eq120}) and
(\ref{eq4}) gives

\begin{eqnarray}
m n_0({\bf r}){\partial^2 \delta {\bf v}\over \partial t^2} &=&
{5\over 3} {\boldmath\nabla} [{\boldmath\nabla}\cdot (P_0({\bf r})
 \delta {\bf v})]
+{2\over 3}{\boldmath\nabla}[n_0({\bf r}) \delta{\bf v}\cdot
{\boldmath\nabla} U_0({\bf r})]\nonumber\\
&+& {\boldmath\nabla}\cdot[n_0({\bf r}) \delta{\bf v}]
{\boldmath\nabla} U_0({\bf r}) \ \ ,
\label{eq121}
\end{eqnarray}
where $\delta{\bf v}\equiv \delta{\bf v}({\bf r},t)$.
Using (\ref{eq7}) to rewrite ${\boldmath\nabla} U_0({\bf r})$ and the fact
that ${\boldmath\nabla} n_0({\bf r}) \propto {\boldmath\nabla} U_0({\bf r})$,
it is straightforward to reduce (\ref{eq121}) to 

\begin{eqnarray}
m {\partial^2 \delta {\bf v}\over \partial t^2} &=&
{5\over 3}{P_0({\bf r}) \over n_0({\bf r})} {\boldmath\nabla}
[{\boldmath\nabla}\cdot\delta {\bf v}]
-{\boldmath\nabla}[\delta{\bf v}\cdot
{\boldmath\nabla} U_0({\bf r})]\nonumber\\
&-& {2\over 3}[{\boldmath\nabla}\cdot\delta{\bf v}]
{\boldmath\nabla} U_0({\bf r}) \ \ .
\label{eq122}
\end{eqnarray}
This is the key result of this letter.
We note that the equilibrium properties of the non-uniform Bose gas only
enter into (\ref{eq122}) through the ratio

\begin{equation}
{P_0({\bf r})\over n_0({\bf r})} =k_BT_0 {g_{5/2}(z_0)\over
g_{3/2}(z_0)}\
,
\label{eq31}
\end{equation}
where the local fugacity $z_0({\bf r})$ has been already defined.

In the absence of a trapping potential we have a uniform gas
and (\ref{eq122}) has the plane wave solution
$\delta{\bf v}({\bf r},t) = \delta{\bf v}_{\omega}({\bf k})
e^{i({\bf k}\cdot{\bf r} - \omega t)}$ with $\delta{\bf v}_{\omega}({\bf k})
\sim {\hat{\bf k}}$.
This is a longitudinal hydrodynamic sound wave with the dispersion relation
$\omega^2 = c^2 k^2$, where the sound velocity is given by

\begin{equation}
c^2 = {5\over 3m}{P_0\over n_0} = {5\over 3} {k_BT_0\over m}{g_{5/2}(z_0) 
\over g_{3/2}(z_0)}~~.
\label{eq126}
\end{equation}
In the classical limit, where $z_0=e^{\mu/k_BT_0} \to 0$,
(\ref{eq126}) reduces to the well known Laplace result \cite{ref8}
$c^2_{cl} = {5\over3}
{k_BT_0\over m}$. At $T_{BEC}$, we have $z_0=1$ since $\mu=0$ and then
$c^2=0.51c^2_{cl}$.

We note that (\ref{eq122}) is a vector equation, giving
three coupled equations for the three components of the velocity
$\delta{\bf v}({\bf r},t)$.
In the presence of the external trap, the normal mode
solutions of this hydrodynamic equation 
are not automatically irrotational and may contain rotational components.
This is a consequence of the inhomogeneous nature of the system.

Special and important solutions of the form $\delta{\bf v}({\bf r},t) 
=\delta{\bf v}_{\omega}({\bf r}) e^{-i\omega t}$ can be obtained 
which are simultaneously irrotational and divergence free.
For these normal modes, (\ref{eq122}) reduces to

\begin{equation}
-m\omega^2 \delta{\bf v}_{\omega}({\bf r}) = - {\boldmath\nabla}
[\delta{\bf v}_{\omega}({\bf r})\cdot{\boldmath\nabla}U_0] \\ .
\label{eq127}
\end{equation}
Since the first term in (\ref{eq122}) makes no contribution,
these solutions are the same in a classical gas as in a highly degenerate
Bose gas just above $T_{BEC}$. These zero-divergence solutions can be shown to
involve isothermal oscillations as follows. Using (\ref{eq9}) and (\ref{eq10}),
one finds ($\Theta\equiv k_BT$)

\begin{eqnarray}
{\partial P\over \partial t} &=&
{1\over \Theta_0} \left({5\over 2}P_0 - n_0\mu_0 \right)
{\partial \Theta \over \partial t} + n_0{\partial \mu \over \partial t}
\label{eq128}\\
{\partial n\over \partial t} &=& {1\over \Theta_0} \left({3\over 2}n_0 -
\gamma_0\mu_0\right)
{\partial \Theta \over \partial t} + \gamma_0{\partial \mu \over \partial t}
\label{eq129} \ .
\end{eqnarray}
where $\gamma_0 \equiv {1\over \Lambda_0^3 \Theta_0} g_{1/2}(z_0)$.
Combining these with (\ref{eq120}) and (\ref{eq4}), one can eliminate
$\partial \mu / \partial t$ and solve for $\partial \Theta /
\partial t$ to obtain

\begin{equation}
{\partial \over \partial t}\Theta({\bf r},t) = -{2\over 3}\Theta_0
{\boldmath\nabla}\cdot
\delta{\bf v}({\bf r},t) \\ .
\label{eq130}
\end{equation}
This result, valid for all $T\ge T_{BEC}$, shows that the temperature
is constant for divergence-free modes.

For an isotropic trap $U_0({\bf r}) = {1\over2}m\omega_0^2r^2$,
a normal mode solution of (\ref{eq127}) is given by
\begin{equation}
\delta{\bf v}_{\omega}({\bf r}) \sim {\boldmath\nabla}[r^{\ell}Y_{\ell m}]
\label{surface}
\end{equation}
with the dispersion relation $\omega=\sqrt{\ell}\omega_0$
holding for $\ell \ge 1$. Using (\ref{eq4}) one
verifies that the associated density fluctuation $\delta n({\bf r},t)
= \delta n_{\omega}({\bf r}) e^{-i\omega t}$ is given by

\begin{equation}
\delta n_{\omega}({\bf r}) \sim r^{\ell-1}Y_{\ell m}(\theta, \phi)
{\partial n_0({\bf r}) \over \partial r} \\ .
\label{eq131}
\end{equation}
In the classical limit, the static density profile reduces to \cite{ref9}

\begin{equation}
n_0({\bf r}) = n_0({\bf r}=0)e^{-m\omega^2_0r^2/2\Theta_0}
\label{eq132}
\end{equation}
and the density fluctuation is then given by
$\delta n_{\omega}({\bf r}) \sim r^{\ell} Y_{\ell m}(\theta, \phi)
n_0(r)$. The latter vanishes at the origin and at
large $r$, being peaked at $r=(\ell\Theta_0/m\omega_0^2)^{1/2}$.
Such zero-divergence modes are usually referred to as surface modes.
It is worth noting that the dispersion relation
of these hydrodynamic surface modes
is identical to the one found by Stringari \cite{ref3} in an
interacting Bose-condensed gas at $T=0$  when the 
equations of motion for the two-component order parameter take the 
form of the equations of hydrodynamics \cite{comment1b}. 
The equation for the velocity fluctuations in Ref.~\cite{ref3}
can be shown to take the simple form

\begin{eqnarray}
m {\partial^2 \delta {\bf v}\over \partial t^2} &=&
-{\boldmath\nabla}
[\delta {\bf v}\cdot {\boldmath\nabla}U_0({\bf r})]
+{\boldmath\nabla}
[(\mu-U_0({\bf r})){\boldmath\nabla}\cdot\delta {\bf v}],
\label{sandro}
\end{eqnarray}
which reduces to (\ref{eq127}) for zero-divergence modes.
This exact correspondence is a surprise, but  only at first sight. 
In fact, these surface modes are entirely
driven by the external force and are insensitive to the form of
the equation-of-state, which is quite different in
the two cases \cite{comment2}. The dispersion relation
of the surface modes discussed above differs from the excitations of
an harmonic oscillator model in the absence
of interactions, which are described by
$\omega=\ell\omega_0$.

We next discuss the compression mode solutions of (\ref{eq122}). The solution
for the ``breathing'' monopole oscillation is obtained
by setting (we assume an isotropic harmonic potential)
$\delta {\bf v}_{\omega}({\bf r}) \sim {\bf r}$. Equation (\ref{eq122}) then
yields $\omega = 2 \omega_0$
and the density fluctuation takes the form $\delta n_{\omega}({\bf r}) \sim
(3 - {m\omega^2_0r^2\over \Theta_0})n_0(r)$ in the classical limit
\cite{comment2}. Because
${\boldmath\nabla}\cdot\delta {\bf v}_\omega({\bf r})$ is
a constant, this monopole solution is unaffected by the quantum terms
associated with Bose statistics which enter (\ref{eq122}) only through
the ratio $P_0/n_0$ given by (\ref{eq31}).
The analogous solution for the interacting Bose-condensed gas
\cite{ref3} at $T=0$ (see (\ref{sandro})) has
the dispersion relation $\omega = \sqrt5 \omega_0$, but
in the non-interacting
model the frequency is given by $\omega = 2 \omega_0$.
The fact that the monopole frequency coincides in the hydrodynamic regime
and in the non-interacting model is a special feature of
a parabolic well. As first shown by Boltzmann \cite{Boltzmann},
for an isotropic harmonic trap, the monopole oscillation of frequency
$2\omega_0$ in a 
classical gas can be shown to be an exact undamped solution of the full
Boltzmann equation \cite{Cercignani,Uhlenbeck}.

Current Bose gas experiments \cite{ref5a,ref5b} usually involve an
anisotropic magnetic trap with axial symmetry

\begin{equation}
U_0({\bf r}) = {1\over 2}m(\omega_{\bot}^2 s^2 +\omega_z^2
z^2)\quad ; \quad
s^2\equiv x^2+y^2\ .
\label{eq43}
\end{equation}
One easily finds that there are exact solutions
of (\ref{eq127}), with frequencies and associated density fluctuations
(for a classical gas) given by

\begin{equation}
\omega^2_{l,m=\pm l} = l\omega_{\bot}^2\quad ; \quad
\delta n_{\omega}({\bf r}) \sim s^{\ell}e^{\pm i\ell \phi}n_0({\bf r})
\end{equation}
\begin{equation}
\omega^2_{l,\pm (l-1)} = (l-1)\omega_{\bot}^2 +
\omega_z^2~ ;~ \delta n_{\omega}({\bf r}) \sim zs^{\ell -1}
e^{\pm i(\ell-1) \phi}n_0({\bf r}).
\label{eq45}
\end{equation}
where $n_0({\bf r})$ is the anisotropic version of (\ref{eq132})
using (\ref{eq43}).  Identical
mode frequencies were obtained by Stringari \cite{ref3}
for an anisotropic condensate at $T=0$, as described by (\ref{sandro}).

In a deformed trap, the monopole mode is coupled
to the quadrupole surface mode ($\ell=2, m=0$). The decoupling is
easily carried out starting from (\ref{eq122}) and
looking for irrotational solutions of the form 
$\delta {\bf v}_\omega({\bf r})
 = \nabla\chi$, with $\chi = \alpha s^2 + \beta z^2$. This gives two coupled
equations for $\alpha$ and $\beta$, which
yield the dispersion relation
\begin{eqnarray}
\omega^2 = {1\over 3}[5 \omega_{\perp}^2 + 4 \omega_z^2
\pm \left(25\omega_{\perp}^4 + 16 \omega_z^4-32 \omega^2_{\perp}\omega^2_z
\right)^{1/2}].
\label{eqcoupling}
\end{eqnarray}
For an isotropic trap, this expression reduces to
$\omega = 2\omega_0$ and $\sqrt2 \omega_0$,
corresponding to the monopole and quadrupole excitations, respectively.
In a cigar-type configuration with $\omega_z \ll \omega_{\perp}$,
the lowest solution of (\ref{eqcoupling}) has the frequency $\omega^2 =
{12 \over 5} \omega_z^2$,
very close to the analogous result $\omega^2 =  {5 \over 2} \omega_z^2$
for the oscillation of the Bose condensate at $T=0$ \cite{ref3}.
This result is consistent with the
recent data of the MIT group \cite{ref5b}.
They found essentially the same normal mode frequencies $\sim 30$Hz at 
very low temperatures ($T\ll T_{BEC}$) and  at $T=2\ T_{BEC}$, the
mode in the latter case being in the hydrodynamic region
since $\nu_c\sim4\times10^3$Hz. In spite of the very similar
frequencies, the damping of these collisionless and hydrodynamic
modes is expected to be quite different \cite{ref1}.

In this paper, we have shown that the complete
spectrum of hydrodynamic normal modes in a trapped Bose gas above
the BEC transition can be obtained from the
solutions of the vector equation in (\ref{eq122}). We 
have used this equation to discuss the normal
modes for which $\nabla\cdot\delta{\bf v}_\omega({\bf r})$
is constant, in which case the first term of the r.h.s. of
(\ref{eq122}) makes no contribution. This is the reason
why the solutions are equally valid for a Bose gas just above $T_{BEC}$
as for a classical gas. We exhibited interesting similarities to
the excitations of a Bose-condensed gas at $T=0$ and of a non-interacting 
gas. We hope that our results stimulate further experimental studies of the 
hydrodynamic modes in trapped Bose gases above the BEC transition.  
There are additional normal mode solutions of (\ref{eq122}) in which 
the first term involving $P_0/n_0$ does contribute. In this regard,
we note that a plot of $g_{5/2}(z_0)/g_{3/2}(z_0)$ in 
(\ref{eq31}) as a function of the distance $r$
from the center of an isotropic parabolic well shows that
this ratio is remarkably close to the classical
value of unity down to about $2T_{BEC}$.
At lower temperatures, it is only slightly less than
unity at small values of $r$. Thus
the solutions of (\ref{eq122}) which depend on the first term can be
expected to be well approximated by the classical limit. These
will be discussed elsewhere.

In a separate publication, we will report on an
extension of the present work to the Bose-condensed region just
below $T_{BEC}$.  This generalized version of
(\ref{eq4})-(\ref{eq6}) leads to the analogue of the two-fluid
superfluid equations \cite{ref17,ref18} for a trapped Bose gas,  with a new
degree of freedom associated with the superfluid component.  The
hydrodynamic modes (first and second sound) of a {\it uniform} weakly
interacting Bose gas have been discussed in Ref.~\cite{GG}.

After completing this work, we received a preprint from Yu. Kagan,
E.L. Surkov and G.V. Shlyapnikov. Their results, based
on the formalism of scaling transformations, agree with our 
expression (\ref{eqcoupling}) for the quadrupole and monopole excitations.


Stimulating discussions with Lev Pitaevskii and Eugene Zaremba are 
acknowledged. A.G. was supported by a grant from NSERC of Canada.


\begin{references}
\bibitem{comment1b} The Thomas-Fermi approximation used at $T=0$ is
only valid if $Na/a_{HO} \gg 1$ \protect\cite{ref3}.
\bibitem{comment1} At finite temperatures, the thermal cloud
has a Gaussian profile which has a size
characterized by $R\sim (2k_B T/m\omega_0^2)^{1/2}$.
At temperatures close to $k_BT_{BEC} \sim \hbar\omega_0 N^{1/3}$,
where $\omega_0$ is the oscillator frequency characterizing the confining
potential and $N$ is the number of atoms in the trap,
the condition $R\gg \ell$ can be shown to be equivalent to 
$N^{2/3}(a/a_{HO})^2 \gg 1$, where $a_{H0}=(\hbar/m\omega_0)^{1/2}$ 
is the harmonic oscillator length.
\bibitem{comment2} Though the velocity fluctuation and the dispersion 
relation in the two regimes coincide, the density fluctuations
do not. This is a result of the different equilibrium density
profiles entering (\protect\ref{eq131}) at $T=0$ \protect\cite{ref3}
and $T>T_{BEC}$.
\end{references}
\end{document}